\documentclass[a4paper,10pt,twoside]{cpc-hepnp}
\usepackage{upgreek,fancyhdr}
\usepackage{multicol}
\usepackage{graphicx}
\usepackage{booktabs}
\usepackage{slashed}
\usepackage{amssymb,bm,mathrsfs,bbm,amscd}
\usepackage[tbtags]{amsmath}
\usepackage{lastpage}
\usepackage{multirow}
\usepackage{textcomp}

\def\leq{\leqslant}
\def\geq{\geqslant}
\DeclareSymbolFont{lettersA}{U}{txmia}{m}{it}
\DeclareMathSymbol{\piup}{\mathord}{lettersA}{25}
\def\hml{\end{multicols}
\end{document}}

\def\BES {\text{BES{\ }\uppercase\expandafter{\romannumeral3}}}
\def\BTWO {\text{Belle{\ }\uppercase\expandafter{\romannumeral2}}}

\usepackage{footnote}
\makesavenoteenv{tabular}

\begin{document}

\newcommand{\calL}{\ensuremath{\mathcal{L}}}
\newcommand{\D}{\displaystyle}
\newcommand{\DF}[2]{\frac{\D#1}{\D#2}}
\newcommand{\be}{\begin{equation}}
\newcommand{\ee}{\end{equation}}
\newcommand{\bea}{\begin{eqnarray}}
\newcommand{\eea}{\end{eqnarray}}

\newcommand{\dq}{{\mathbf q} }

\fancyhead[c]{\small Chinese Physics C~~~Vol. 41, No. 5 (2017)
053001} \fancyfoot[C]{\small 053001-\thepage}

\footnotetext[0]{Received 24 October 2016, Revised 17 January
2017}

\title{Physics potential of searching for 0$\upnu\upbeta\upbeta$ decays in JUNO\thanks{Supported by
Strategic Priority Research Program of Chinese Academy of
Sciences(XDA10010900), CAS Center for Excellence in Particle
Physics (CCEPP), Postdoctoral Science Foundation of China and
Chinese Academy of Sciences (2015IHEPBSH101), Program of
International S\&T Cooperation of MoST (2015DFG02000)}}

\author{Jie Zhao$^{1;1)}$\email{zhaojie@ihep.ac.cn}%
\quad Liang-Jian Wen$^{1,2;2)}$\email{wenlj@ihep.ac.cn}%
\quad Yi-Fang Wang$^{1,2}$
\quad Jun Cao$^{1,2}$
}\maketitle

\address{$^1$ Institute of High Energy Physics, Chinese Academy of Sciences, Beijing 100049, China\\
$2$ State Key Laboratory of Particle Detection and Electronics
(Institute of High Energy Physics, Chinese Academy of\\ Sciences
and University of Science and Technology of China)}

\begin{abstract}
In the past few decades, numerous searches have been made for the
neutrinoless double-beta decay (0$\upnu\upbeta\upbeta$) process,
aiming to establish whether neutrinos are their own antiparticles
(Majorana neutrinos), but no 0$\upnu\upbeta\upbeta$ decay signal
has yet been observed. A number of new experiments are proposed
but they ultimately suffer from a common problem: the sensitivity
may not increase indefinitely with the target mass. We have
performed a detailed analysis of the physics potential by using
the Jiangmen Underground Neutrino Observatory (JUNO) to improve
the sensitivity to 0$\upnu\upbeta\upbeta$ up to a few meV, a major
step forward with respect to the experiments currently being
planned. JUNO is a 20 kton low-background liquid scintillator (LS)
detector with 3\%/$\sqrt{E \text{(MeV)}}$ energy resolution, now
under construction. It is feasible to build a balloon filled with
enriched xenon gas (with $^{136}$Xe up to 80\%) dissolved in LS,
inserted into the central region of the JUNO LS. The energy
resolution is $\sim$1.9\% at the $Q$-value of $^{136}$Xe
0$\upnu\upbeta\upbeta$ decay. Ultra-low background is the key for
0$\upnu\upbeta\upbeta$ decay searches. Detailed studies of
background rates from intrinsic 2$\upnu\upbeta\upbeta$ and $^{8}$B
solar neutrinos, natural radioactivity, and cosmogenic
radionuclides (including light isotopes and $^{137}$Xe) were
performed and several muon veto schemes were developed. We find
that JUNO has the potential to reach a sensitivity (at 90\% C. L.)
to $T^{0\upnu\upbeta\upbeta}_{1/2}$ of $1.8\times10^{28}$ yr
($5.6\times10^{27}$ yr) with $\sim$50 tons (5 tons) of fiducial
$^{136}$Xe and 5 years exposure, while in the 50-ton case the
corresponding sensitivity to the effective neutrino mass,
$m_{\upbeta\upbeta}$, could reach (5--12) meV, covering completely
the allowed region of inverted neutrino mass ordering.
\end{abstract}


\begin{pacs}
23.40.-s, 14.60.Pq, 21.10.Tg\qquad     {\bf DOI:}
10.1088/1674-1137/41/5/053001
\end{pacs}


\footnotetext[0]{\hspace*{-6mm}\raisebox{-1.05ex}{\includegraphics[scale=0.35]{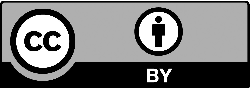}}
Content from this work may be used under the terms of the Creative
Commons Attribution 3.0 licence. Any further distribution of this
work must maintain attribution to the author(s) and the title of
the work, journal citation and DOI. Article funded by SCOAP$^3$
and published under licence by Chinese Physical Society and the
Institute of High Energy Physics of the Chinese Academy of
Sciences and the Institute of Modern Physics of the Chinese Academy of Sciences and IOP Publishing Ltd}%

\begin{multicols}{2}

\section{Introduction}

Currently, the neutrinoless double-beta decay
(0$\upnu\upbeta\upbeta$) process is the only experimentally
feasible and most sensitive way to probe if massive neutrinos are
their own antiparticles, namely, Majorana particles. Violation of
lepton number is a direct consequence of the
0$\upnu\upbeta\upbeta$ process, in which a nucleus decays by
emitting two electrons and nothing else, $N(A,Z)\rightarrow N(A,
Z+2) + 2{\rm e}^-$. Furthermore, searching for
0$\upnu\upbeta\upbeta$ decay can shed light on the absolute scale
of neutrino masses.

Under the standard three neutrino framework, the effective
neutrino mass in 0$\upnu\upbeta\upbeta$ decay is defined as
$m_{\upbeta\upbeta}\equiv|\sum_i (m_i U^2_{{\rm e}i})|$, where
$U_{{\rm e}i}$ (for $i=1,\,2,\,3$) denote the matrix elements in
the first row of lepton flavor mixing matrix $U$, and $m_i$ (for
$i=1,\,2,\,3$) are neutrino masses. For Majorana neutrinos,
$m_{\upbeta\upbeta}$ is sensitive to the neutrino masses, neutrino
mixing angles and Majorana $CP$ phases. Using the standard
parametrization of $U$, $m_{\upbeta\upbeta}=|m_1 c^2_{12} c^2_{13}
{\rm e}^{2{\rm i}\phi_1} + m_2 s^2_{12} c^2_{13} {\rm e}^{2{\rm
i}\phi_2} + m_3 s^2_{13}|$ ~\cite{Zhang:2015kaa}, where
$c_{12}=\cos\theta_{12}$, $c_{13}=\cos\theta_{13}$,
$s_{13}=\sin\theta_{13}$, $s_{12}=\sin\theta_{12}$,
$\{\theta_{12},\theta_{13}\}$ are neutrino mixing angles, and
$\{\phi_1, \phi_2\}$ are Majorana $CP$ phases. In this definition
of $m_{\upbeta\upbeta}$, it has been assumed that the
0$\upnu\upbeta\upbeta$ process is dominated only by the exchange
of light Majorana neutrinos. Sterile neutrinos or other exotic
physics are not considered. If neutrinos have an inverted mass
ordering, $m_{\upbeta\upbeta}$ will be greater than $\sim$0.015
eV, based on current and projected knowledge of the neutrino
mixing parameters~\cite{An:2015jdp}. For the normal neutrino mass
ordering case, no lower bound exists and $m_{\upbeta\upbeta}$
could vanish due to the cancellation among the $m_iU^2_{{\rm e}i}$
terms that are modulated by the Majorana\vspace{-4.3mm}\linebreak

\end{multicols}
\begin{multicols}{2}
\noindent
phases.

Enormous experimental efforts have been made to search for
0$\upnu\upbeta\upbeta$ in the last few decades, using various
nuclear isotopes, such as $^{136}$Xe, $^{76}$Ge, $^{130}$Te, etc,
as discussed in recent reviews (see~\cite{Ostrovskiy:2016uyx} and
references therein). None of them observed a
0$\upnu\upbeta\upbeta$ decay signal. It is desirable  for the next
generation 0$\upnu\upbeta\upbeta$ experiments to have a
sensitivity of $m_{\upbeta\upbeta}\sim$10 meV. With such
sensitivity, if the neutrino mass ordering is determined to be
inverted by future reactor and accelerator experiments, either a
positive (observation of 0$\upnu\upbeta\upbeta$ decay) or a
negative (no observation) result would be able to probe the
Majorana nature or Dirac nature of neutrinos, respectively.

The half-life of 0$\upnu\upbeta\upbeta$, $T^{0\upnu}_{1/2}$, is
related to the effective Majorana neutrino mass,
$m_{\upbeta\upbeta}$, by a phase space factor $G_{0\upnu}$ and a
nuclear matrix element (NME)
$\mathcal{M}_{0\upnu}$:\vspace{-1.2mm}
\begin{equation}
  (T^{0\upnu}_{1/2})^{-1} = G_{0\upnu} |\mathcal{M}_{0\upnu}|^2 m^2_{\upbeta\upbeta}\vspace{-1.2mm}
\end{equation}
where both  $G_{0\upnu}$ and $\mathcal{M}_{0\upnu}$ can be
calculated theoretically. However, the NME has relatively large
uncertainties from different nuclear models, see
Ref.~\cite{Engel:2015wha} and references therein.

Two-neutrino double-beta decay (2$\upnu\upbeta\upbeta$) is allowed
by the Standard Model and has been observed in many nuclei.
0$\upnu\upbeta\upbeta$ can be distinguished from
2$\upnu\upbeta\upbeta$ by measuring the sum energy of the two
electrons and looking for a mono-energetic peak at the $Q$-value.
The region around the $Q$-value is referred to as the
0$\upnu\upbeta\upbeta$ window, namely the region of interest
(ROI). Different experiments\vspace{1.3mm} might choose different
ROIs, e.g, $\pm1\sigma$, $\pm2\sigma$, $\pm\dfrac{1}{2}$FWHM
\vspace{1.3mm}or even an asymmetric window around the $Q$-value,
due to different background levels and energy resolutions.
Excellent energy resolution and ultra-low background in the ROI
are the keys to searching for 0$\upnu\upbeta\upbeta$.

The Jiangmen Underground Neutrino Observatory (JUNO) is a
multi-purpose experiment that primarily aims to determine the
neutrino mass ordering and to measure precisely the neutrino
mixing parameters~\cite{An:2015jdp,Djurcic:2015vqa}. Such
precision measurement could reduce the range of $T^{0\upnu}_{1/2}$
predictions by a factor of 2~\cite{Ge:2015bfa}. The way to
distinguish the neutrino mass ordering at JUNO is via exploring
the effect of interference between atmospheric- and solar-$\Delta
m^2$ driven
oscillations~\cite{Zhan:2008id,Zhan:2009rs,Li:2013zyd}. The
baseline design of the JUNO detector is a 20 kton low-background
liquid scintillator (LS) with an unprecedented energy resolution
($\sigma/E$) of 3\%/$\sqrt{E \text{(MeV)}}$. At the Q-value of
$^{136}$Xe 0$\upnu\upbeta\upbeta$ decay
($Q^{0\upnu\upbeta\upbeta}$=2457.8 keV), or $^{130}$Te
($Q^{0\upnu\upbeta\upbeta}$=2530 keV), the energy resolution is
expected to be $\sim$1.9\%, which is suitable for a
$0\upnu\upbeta\upbeta$ search. In addition, online purification is
another advantage of LS detectors, and the liquid can reach the
adequate level for 0$\upnu\upbeta\upbeta$ searches.
KamLAND-Zen~\cite{KamLANDZen:2012aa} and
SNO+~\cite{Andringa:2015tza} are two examples, using $^{136}$Xe
and $^{130}$Te isotopes, respectively. However, their detector
sizes are limited so that their sensitivity to
$m_{\upbeta\upbeta}$ can only reach a few tens of meV. A very
large LS detector can perform a better
measurement~\cite{Jaffe:2015pla}.

In this paper, we explore the physics potential of searching for
0$\upnu\upbeta\upbeta$ decays of $^{136}$Xe with the JUNO
detector, aiming for a few meV sensitivity on
$m_{\upbeta\upbeta}$, by dissolving enriched pure xenon gas into
the liquid scintillator. The Xe-loaded LS target could be
separated from the normal LS by deploying a highly transparent and
clean balloon. The clean normal LS can provide sufficient passive
shielding against external radioactivity, and act as an active
zone to track the muons and veto the cosmogenic backgrounds.

\section{JUNO detector}
\label{sec:detector}

The JUNO site has an overburden of $\sim$ 700 m rock. The central
detector (CD) is an acrylic sphere of 35.4 m in diameter, holding
the 20 kton LS, supported by a spherically latticed shell made of
stainless steel (SS) with a diameter of 40.1 m. About $\sim$18000
20-inch PMTs are mounted on the SS latticed shell, looking inward
towards the LS target. In addition, up to $\sim$36000 3-inch PMTs
will be installed in the gaps between the 20-inch PMTs, to form a
complementary calorimetry system and improve the muon measurement.
Outside the SS latticed shell, an ultra-pure water pool of 43.5 m
diameter and 44 m depth is equipped with $\sim$2000 20-inch PMTs,
providing an active cosmic muon veto as a water Cerenkov detector
and sufficient passive shielding from the environmental
radioactivity. On top of the water pool, the
OPERA~\cite{Acquafredda:2009zz} target trackers are re-used as a
complementary Top Tracker system, providing precise track
measurement of cosmic muons.

The JUNO LS uses linear alkyl-benzene (LAB) as the solvent,
2,5-diphenyloxazole (PPO) as the primary fluor, and
1,4-bis[2-methylstyryl]benzene (bis-MSB) as the wavelength
shifter. The current baseline recipe is adopted from the Daya Bay
experiment~\cite{Ding:2008zzb,Beriguete:2014gua} but without
gadolinium doping. As discussed in~\cite{An:2015jdp}, the baseline
LS purity requirement for reactor antineutrino studies is less
than 10$^{-15}$ g/g for $^{238}$U and $^{232}$Th, 10$^{-16}$ g/g
for $^{40}$K and 1.4$\times10^{-22}$ g/g for $^{210}$Pb. This is
sufficient for the determination of neutrino mass ordering. A
sophisticated on-line purification system can be set up, and
eventually two orders of magnitude better purity is expected to be
achievable. Such optimal purity (10$^{-17}$ g/g for $^{238}$U and
$^{232}$Th, 10$^{-18}$ g/g for $^{40}$K and $10^{-24}$ g/g for
$^{210}$Pb) is adequate for 0$\upnu\upbeta\upbeta$ searches. The
backgrounds caused by the internal impurities are discussed in
Section~3.3.

The target element for 0$\upnu\upbeta\upbeta$ searches in this
study, as an example, is chosen to be $^{136}$Xe for its high
purity, high $Q$-value, and high solubility in LS. Of course other
elements are not excluded at present. $^{130}$Te is another
possible element and has a natural abundance of 34.1\%. It is
technically challenging to purify tellurium and reach
\textgreater5\% tellurium loading in LAB-based scintillator. As an
example, in the Te-loaded phase of the SNO+ experiment, with 0.3\%
Te-loading, the projected $^{238}$U and $^{232}$Th concentration
would be two orders of magnitude worse than the pure LAB-PPO
scintillator~\cite{Andringa:2015tza}. The stability, transparency
and light yield would also decrease with high tellurium loading.
Unlike xenon, cosmogenic activation of the tellurium nuclei could
produce a large number of long-lived radioactive isotopes. To
suppress such background, the exposure time of tellurium on the
surface should be controlled. A purification process and
additional long cooling time underground is
necessary~\cite{Lozza:2014haa}. At the depth of JUNO, the
cosmogenic background could be serious for 0$\upnu\upbeta\upbeta$
searches. In this study, we choose $^{136}$Xe as an example to
evaluate the physics potential of the 0$\upnu\upbeta\upbeta$
search at JUNO. The possibility of using $^{130}$Te will be
evaluated in future.

A transparent and strong balloon can be used to separate the Xe-LS
from the normal LS. Xenon gas is found to be soluble into liquid
scintillator more than 3\% by weight, but the light yield could be
reduced depending on the xenon concentration~\cite{Gando2011}. We
expect that such an effect can be compensated by tuning the
concentration of the fluors. Thus we assumed 5\% by weight of the
enriched xenon gas ($^{{\rm enr}}$Xe) that consists of 80\%
$^{136}$Xe. \vspace{1.5mm}We chose the ROI as the
$\pm\dfrac{1}{2}$FWHM region around the
\vspace{1.5mm}$Q_{\upbeta\upbeta}$ value. The parameters that were
chosen in our calculation are compared with the KamLAND-Zen
detector in Table~1. The efficiency of 0$\upnu\upbeta\upbeta$
events in the ROI, defined as
$\varepsilon_{0\upnu\upbeta\upbeta}$, was calculated according to
the energy resolution at $Q_{\upbeta\upbeta}$ and the selected ROI
window.

\begin{center}
\tabcaption{Comparison of the parameters of the assumed JUNO Xe-LS
detector and KamLAND-Zen detector.} \footnotesize
\begin{tabular*}{86mm}{c@{\extracolsep{\fill}}ccc}
\toprule & KamLAND-Zen & JUNO Xe-LS \\
\hline
energy resolution & 6.6\%/$\sqrt{E}$~\cite{KamLANDZen:2012aa} & 3\%/$\sqrt{E}$ \\
 & 7.3\%/$\sqrt{E}$~\cite{KamLAND-Zen:2016pfg} &  \\
Xe-doping & 2.5\% (phase I~\cite{KamLANDZen:2012aa}) & 5\% \\
 & 2.9\% (phase II~\cite{KamLAND-Zen:2016pfg}) &  \\
$^{136}$Xe enrichment & $\sim$91\%~\cite{KamLANDZen:2012aa,KamLAND-Zen:2016pfg} & 80\% \\
$0\upnu\upbeta\upbeta$ ROI & (2.3, 2.7) MeV~\cite{KamLAND-Zen:2016pfg} & (2403, 2513) keV\\
$\varepsilon_{0\upnu\upbeta\upbeta}$ in ROI &
89.9\%\footnote{corresponding to $\sigma\sim$7.3\%$\sqrt{E}$
resolution} & 75.8\% \\ \bottomrule
\end{tabular*}
\end{center}\vspace{-1mm}
\noindent {\footnotesize

${}^*$corresponding to $\sigma\sim7.3$\%$\sqrt{E}$ resolution}

\section{Backgrounds}

The natural radioactivity in the liquid scintillator and the
long-lived radioactive isotopes produced by muon spallation are
the dominant background for the $0\upnu\upbeta\upbeta$ search. The
spallation neutrons produced by cosmic muons can induce the
$\upbeta$-decay isotope $^{137}$Xe, with a half-life of 3.82
minutes, via the $^{137}$Xe$({\rm n},\gamma)$ reaction. The
$Q$-value for $^{137}$Xe decay is 4173$\pm$7
keV~\cite{Browne20072173}, so the $\upbeta$ spectrum overlaps the
$Q$-value of $^{136}$Xe $0\upnu\upbeta\upbeta$ decay. The
background rates are evaluated below.

\subsection{Intrinsic 2$\upnu\upbeta\upbeta$ background}

With finite energy resolution, 2$\upnu\upbeta\upbeta$ events
leaking into the 0$\upnu\upbeta\upbeta$ ROI are the intrinsic
background. Such background decreases dramatically as energy
resolution improves. Hereafter, the background index, defined as
the background rate per unit $^{136}$Xe mass per ROI, was
introduced to quantify the background. We estimated the intrinsic
2$\upnu\upbeta\upbeta$ background rate to be 0.2/ROI/(ton
$^{136}$Xe)/yr by convoluting the theoretical
2$\upnu\upbeta\upbeta$ energy spectrum~\cite{Schenter:Vogel} with
the detector energy resolution curve.

\subsection{Solar-$\upnu$ background}

The $\upnu$-e scattering signal from $^{8}$B solar neutrinos has a
continuous spectrum up to \textgreater10 MeV, thus it can also
contribute to the ROI background. Its signal rate was estimated to
be 4.5/kton/day~\cite{An:2015jdp}. Using the simulated energy
spectrum of the $^{8}$B $\upnu$-e scattering signal, also
described in~\cite{An:2015jdp}, we estimated the background index
to be 28/ROI/(kton Xe-LS)/yr, equivalent to 0.7/ROI/(ton
$^{136}$Xe)/yr under the assumption of 5\% $^\text{enr}$Xe.

If natural xenon gas is used instead of $^{136}$Xe-enriched xenon
gas, the background index from the solar neutrinos would be 10
times larger, since the $^{136}$Xe abundance in natural xenon is
only $\sim$8\%.

\subsection{Natural radioactivity}
\label{sec:subsection:internal}

\subsubsection{Internal $^{238}$U and $^{232}$Th contamination}

The projected radioactivities of the JUNO detector components such
as liquid scintillator, PMT glass, acrylic and supporting
structures were discussed in~\cite{An:2015jdp,Li:2015cqa}. The
external radioactivities could be eliminated by a sufficient
fiducial volume cut, e.g, 1 m inward from the LS edge, thus only
the internal LS radio-impurities need to be considered. As
discussed in Section~2, an optimal radio-purity level
$\sim10^{-17}$ g/g for U and Th is reachable. The following
studies are based on this optimal radio-purity assumption.

The $\upbeta+\upgamma$ emissions from $^{214}$Bi ($^{238}$U chain,
$Q=3.272$ MeV) could be a serious background for
0$\upnu\upbeta\upbeta$ searches, because there is a 2.448 MeV
$\upgamma$ line, which can leak into the ROI. From the simulated
energy spectra of events from the $^{238}$U chain, the background
index was calculated to be 8.3/ROI/(ton $^{136}$Xe)/yr. The
$^{214}$Bi-$^{214}$Po $\upbeta$-$\upalpha$ cascade decay
($\tau=237~\upmu$s) is very effective at rejecting $^{214}$Bi
events. The $\alpha$ energy from $^{214}$Po decay is 7.686 MeV,
and its quenched response is well above the detector threshold,
resulting in a high efficiency of tagging $^{214}$Bi events in the
ROI. We evaluated the background rejection with Monte Carlo
samples by requiring the time and distance between the prompt
$\upbeta$ and delayed $\upalpha$ decay events to be less than 2.0
ms and 2.0 m, respectively. The residual background is due to the
Bi-Po cascade decays that have a decay time longer than 2.0 ms, or
occurred within one readout window (nominally 1 $\upmu$s for JUNO)
and their summed energy falls into the ROI. We found $\sim$99.97\%
of the events in the ROI from $^{238}$U chain were rejected,
resulting in \textless0.003/ROI/(ton $^{136}$Xe)/yr residual ROI
background.

The fast $^{212}$Bi-$^{212}$Po $\upbeta$-$\upalpha$ cascade decay
from the $^{232}$Th chain ($\tau=431$ ns) leads to 90\% of the two
signals occurring in the 1$\mu$s nominal readout window. Our
GEANT4 MC indicated that the summation of the visible energies of
$^{212}$Bi $\upbeta+\upgamma$ ($Q=2.252$ MeV) and $^{212}$Po
$\alpha$ ($Q=8.954$ MeV) had a fraction of 6.2\% inside the ROI
window, while neither the individual $\upbeta$ nor $\alpha$ decays
could contribute to the ROI. Assuming $10^{-17}$ g/g $^{232}$Th
concentration, we estimated the background index from the
summation events to be 1.25/ROI/(ton Xe-LS)/yr. Thus, special care
should be taken to distinguish and reject these two decays. JUNO
will adopt 1 GHz Flash ADC (FADC) to record the full waveforms
from all the PMTs inside the readout window, allowing a pulse
shape discrimination (PSD) approach to distinguish two decays
which are close in time. The LAB-based liquid scintillator was
demonstrated to have good capability of e$^-$/$\upalpha$
discrimination~\cite{1674-1137-35-11-009}. A full MC simulation
including scintillation processes and PMT timing resolution was
performed for the decays. We developed a PSD method by using the
width and the tail fraction of the measured scintillation time
profile, in which the time-of-light of photons were corrected. The
discrimination efficiency was found to reach \textgreater97.5\%,
resulting in a residual ROI background of 0.03/ROI/(ton
$^{136}$Xe)/yr. Our GEANT4 MC indicated negligible contribution
from internal $^{208}$Tl decays ($Q=4.999$ MeV) to the ROI,
because the visible energy inside the LS is the summation of the
$\upbeta$ and $\upgamma$ energies, which has a minimum energy of
3.2 MeV. This is different from the surface contamination, where
$\beta$s deposit their energy in the vessel material without
scintillation, but the 2.615 MeV $\upgamma$s could leak into the
ROI.

For comparison, the $^{238}$U and $^{232}$Th contamination in
Borexino LS detector reached $<$10$^{-18}$
g/g~\cite{Bellini:2014uqa}, while the radio-purity in KamLAND-Zen
detector is $\sim3.5\times10^{-16}$ g/g for $^{238}$U and
$\sim2.2\times10^{-15}$ g/g for $^{232}$Th,
respectively~\cite{KamLANDZen:2012aa}. Past experiences in LS
purification would benefit JUNO to reach its radio-purity goal.

\subsubsection{External radioactivity}
\label{sec:subsection:extRad}

As discussed in Section~2, a highly transparent balloon can be
used to contain the Xe-LS. Although the balloon material could be
very radio-pure (e.g, ppt level), the possible dust contamination
during installation and the radon contamination during LS
purification could yield much higher $^{214}$Bi levels on the
surface of the balloon. A fiducial volume cut is effective against
$^{214}$Bi and $^{208}$Tl decays from the balloon. We consider
that a 1 m cut from the Xe-LS target edge would be sufficient.

Extreme care should be taken to prevent radon (mainly $^{222}$Rn ,
$\tau=5.52$ day) from penetrating into the Xe-LS during the
purification process. We put a requirement of 5$\times10^3$
atoms/(kton Xe-LS)/yr external radon leakage rate. Taking into
account the 99.97\% rejection efficiency via $^{214}$Bi-$^{214}$Po
tagging, it would lead to a 0.2/ROI/(ton $^{136}$Xe)/yr background
rate.

\subsection{Cosmogenic backgrounds}
\label{sec:subsection:cosmo}

Energetic cosmic muons can cause spallation in organic liquid
scintillator, and produce long-lived radioactive isotopes via the
photon-nuclear or hadronic processes. The overburden for the JUNO
detector is 748 m, and the muon flux at the JUNO site is about
0.003 Hz/m$^2$, which is a factor of $\sim$2 more than the
underground lab at Kamioka. The rate of muons passing through the
JUNO LS volume is about 3.0 Hz, with a mean energy of 215 GeV.

\end{multicols}
\begin{center}
\tabcaption{Summary of the simulated muon-induced radioactive
isotopes (mostly with $Z\leq6$) in the JUNO LS. Only the isotopes
that can contribute to the 0$\upnu\upbeta\upbeta$ window are
listed. $^{10}$C, $^{6}$He, $^{8}$Li and $^{12}$B are the four
dominant contributors, while the contributions from other isotopes
are combined, such as $^{11}$Be ($\tau_{1/2}=$13.8 s,
$Q(\upbeta^-)=11.5$ MeV), $^{9}$C ($\tau_{1/2}=0.13$ s,
$Q(\upbeta^+)=16.5$ MeV), $^{16}$N ($\tau_{1/2}=7.13$ s,
$Q(\upbeta^-\upgamma)=10.4$ MeV), $^{9}$Li ($\tau_{1/2}=0.178$ s,
$Q(\upbeta^-\upgamma$-${\rm n})=13.6$ MeV), $^{8}$He
($\tau_{1/2}=0.12$ s, $Q(\upbeta^{-}\upgamma$-${\rm n})=10.7$
MeV). The latter two isotopes are $\upbeta$-${\rm n}$ emitters
with branching ratios of 51\% and 16\%, respectively. Such
$\upbeta$-${\rm n}$ decays can be rejected by coincidence cuts and
were removed in this table. } \footnotesize  \tabcolsep 5.2pt
\begin{tabular*}{175mm}{cccccccccccc}
\toprule  & $T_{1/2}$ in & radiation $E$/ &
\multicolumn{2}{c}{$R_{{\rm prod}}$/(ton$^a\cdot$yr)$^{-1}$} &
primary &
\multicolumn{4}{c}{accompanied neutrons} & background Index \\
\cline{7-10}
 & JUNO LS & MeV &  FLUKA~\cite{An:2015jdp} & this work & process & 0n & 1n & 2n & 3n & /(ROI$\cdot$ton$^b\cdot$yr)$^{-1}$ \\
\hline
$^{10}$C & 19.3 s & 3.65 ($\upbeta^+\gamma$) & 9.8 & 9.3 & $\uppi^+$ Inelastic & 2.20\% & 37.4\% & 38.3\% & 22.1\% & 16.4
\\[-0.3mm]
$^{6}$He & 0.807 s &  3.51 ($\upbeta^-$)  & 11.0 & 6.1 & n Inelastic & 39\% & 42.7\% & 10.5\% & 6.7\% & 8.8-4.9 \\[-0.5mm]
$^{8}$Li & 0.84 s&  16.0 ($\upbeta^-\upalpha$)  & 19.0 & 8.4 & n Inelastic & 62.8\% & 19.2\% & 16.3\% & 1.5\% & 3.4-1.5\\[-0.5mm]
$^{12}$B & 0.02 s & 13.4 ($\upbeta^-$)  & 19.6 & 12.4 & n Inelastic & 93.6\% & 6.3\% & $<$0.1\% & --- & 3.0-1.9 \\[-0.5mm]
others & --- & ---  & 2.5 & 0.81 & --- & --- & --- & --- & --- & 0.51 \\
\bottomrule
\end{tabular*}%
\end{center}
\vspace{-0.8mm} \noindent {\footnotesize

${}^a$here ton is a unit of Xe-LS mass

${}^b$here ton is a unit of $^{136}$Xe mass}

\begin{multicols}{2}

The production of the radioactive isotopes in JUNO LS was
evaluated by GEANT4~\cite{Agostinelli:2002hh} simulation. A Monte
Carlo (MC) muon data set of $\sim$342 days' worth of statistics
was produced to study the cosmogenic backgrounds in the
0$\upnu\upbeta\upbeta$ search. The results are summarized in
Table~2, including the raw production rates, the primary
production processes, the fractions for different number of
accompanied neutrons and the background indexes in the ROI. The
production rates from the earlier analysis~\cite{An:2015jdp} using
FLUKA~\cite{Ferrari:2005zk} were also listed for comparison. Both
GEANT4 and FLUKA indicate that $^{10}$C, $^{6}$He, $^{8}$Li and
$^{12}$B are the dominant contributors. Other isotopes were found
to have relatively small contributions in the ROI, thus they were
combined in the last row of the table. Given their long half-lives
and relatively high muon rate in the JUNO detector, it was
challenging to reject those backgrounds. In Table~2, our GEANT4 MC
predicted a similar $^{10}$C production yield to FLUKA, whereas it
gave a lower $^{6}$He, $^{8}$Li and $^{12}$B production yield than
FLUKA. This is probably due to different hadronic interaction
models being used. In the following analysis we used the newly
produced GEANT4 MC data with large statistics, showing that the
residual cosmogenic backgrounds after muon veto is evaluated to be
$\sim$10\% of the total background. Thus the differences between
FLUKA and GEANT4 were considered not to affect the main conclusion
of this paper. To mimic a real data set, we assigned a time stamp
for each primary muon and its daughters according to the average
$R_\mu$=3 Hz muon rate, then the primary muons and their
subsequent events were mixed and sorted.\vspace{0.2mm}

Cosmogenic isotopes are mainly produced by energetic showering
processes in the LS. Table~2 shows that $\sim$98\% of $^{10}$C,
$\sim$60\% of $^{6}$He and $\sim$37\% of $^{8}$Li are accompanied
by $\geq$1 neutrons, allowing us to develop a special veto
strategy to reject those $\upbeta$-decays. Although the $^{12}$B
production has weak correlation with neutrons, it has a relatively
short half-life and thus can be efficiently rejected by vetoing a
longer time. The veto methods to reject the cosmogenic backgrounds
and the results are discussed in following
subsections.\vspace{0.2mm}

The previous measurements~\cite{Abe:2009aa} and
simulations~\cite{Li:2015lxa}, as well as our simulation show that
the distance from the isotope's production position to its parent
muon track approximately follows an exponential profile. Thus,
vetoing a cylindrical volume along the reconstructed muon track
for sufficient time can significantly reduce the muon induced
backgrounds.\vspace{0.2mm}

As described in Section~2, the JUNO central detector will be
equipped with a vast number of 3-inch PMTs, providing excellent
track reconstruction for both minimum ionizing muons and showering
muons. However, the track reconstruction of a showering muon is
non-trivial. Our simulation indicated that a muon changes little
in its direction after producing a shower. Thus the entry and exit
points in the pattern of hit PMTs can give a good estimation of
the muon track. In addition, we found that high multiplicity
neutrons were produced near the high ${\rm d}E/{\rm d}x$ region,
and those neutrons' vertices could be used to further constrain
the muon track and reconstruct the location of the muon shower.

The muon events were first categorized into two types: the normal
muons ($\mu_{{\rm norm}}$) and the neutron-associated muons
($\mu_{{\rm n-assoc}}$). Their identification and corresponding
veto criteria are described below:

1) $\mu_{{\rm norm}}$ identification: if the distance from the LS
center to the muon track is within $(R_{{\rm Xe}}+3)$ meters,
where $R_{{\rm Xe}}$ is the radius of Xe-LS volume. $\mu_{{\rm
norm}}$ veto: any signal within a veto time window of 1.2 s and
within a 3 m cylinder along the muon track was rejected.

2) $\mu_{{\rm n-assoc}}$ identification: among the $\mu_{{\rm
norm}}$ samples, if a neutron-like signal occurs within 1 ms after
the muon and within ($R_{{\rm Xe}}+2$) meters from the detector
center. The neutron-like signal is identified as an event in the
n-capture energy window, (2.0, 2.4) MeV.

$\mu_{{\rm n-assoc}}$ veto: any signal within 2 meters from each
associated neutron-like signal and within a veto time window of
$t^{{\rm veto}}_{n-\mu}$ was rejected.

We evaluated the efficiency of the muon veto and the residual
cosmogenic background for different target radii $R_{Xe}$. For
each assumed $R_{{\rm Xe}}$, the muons were first categorized
according to the above criteria. By definition, the rates of
$\mu_{{\rm norm}}$ and $\mu_{{\rm n-assoc}}$ depend on the Xe-LS
target size $R_{{\rm Xe}}$. With the MC data set, the rates were
parameterized as $R_\mu^{{\rm norm}}=9.38\times10^{-3}\cdot
(R_{{\rm Xe}}+3)^2$ Hz and $R_\mu^{{\rm
n-assoc}}=3.58\times10^{-5}\cdot (R_{{\rm Xe}}+2)^3$ Hz,
respectively. Then we applied the above muon veto strategies to
the mixed MC data set, and particularly tested different values of
the veto window $t^{{\rm veto}}_{n-\mu}$. Finally the live time
and the rate of residual background were calculated.

\subsubsection{Long-lived light isotopes}

Among the dominant isotope backgrounds, $^{10}$C has the longest
half-life $\tau(^{10}{\rm C})=27.8$ s, thus the veto window
$t^{{\rm veto}}_{n-\mu}$ should be sufficiently long to reject
$^{10}$C and $^{6}$He effectively. We tested different $t^{{\rm
veto}}_{n-\mu}$: $2\tau(^{10}{\rm C})$, $4\tau(^{10}{\rm C})$ and
$6\tau(^{10}{\rm C})$, as shown in Table~3.

Increasing $t^{{\rm veto}}_{n-\mu}$ significantly reduced the
$^{10}$C and $^{6}$He rates, with negligible loss of live-time due
to the low rate of $\mu_{{\rm n-assoc}}$. When using the
$\mu_{{\rm norm}}$ veto plus $6\tau(^{10}{\rm C})$ window for
$\mu_{{\rm n-assoc}}$ veto, the reduction factors for $^{10}$C and
$^{6}$He were 309 and 78, respectively. Although the MC indicated
that the $^{12}$B production had a weak correlation with the
neutron production, it was also strongly suppressed after applying
the above muon veto, due to a much shorter half-life. Table~3
showed that with a proper muon veto the cosmogenic backgrounds
could be well controlled.

To estimate the veto efficiency, tracer events that were uniformly
distributed in time and within the LS volume were mixed into the
sorted MC data set. After applying the selections cuts, the
efficiency was estimated as $M_{\rm s}/M$, where $M$ was the total
number of tracer events and $M_{\rm s}$ was the number of tracer
events that survived the veto. The efficiency was precisely
calculated with large statistics of the tracer events. In Table~3,
the efficiency varied a little when adding the $\mu_{{\rm
n-assoc}}$ veto. In addition, we found the efficiency and the
background index slightly changed for different Xe-LS target sizes
$R_{{\rm Xe}}$. The last column in the table was used for
sensitivity calculation.

\begin{center}
\tabcaption{Background indices of the muon-induced radioactive
isotopes, for different muon veto schemes. The values are based on
GEANT4 simulation. $^{10}$C, $^{6}$He, $^{8}$Li and $^{12}$B are
the four dominant contributors to the ROI background, while the
contribution from other isotopes are combined. Taking the FLUKA
results from Table~2, the total residual background would increase
to 0.21/ROI/(ton $^{136}$Xe)/yr after $\mu_{{\rm norm}}$ veto plus
$\mu_{{\rm n-assoc}}$ veto with $t^{{\rm
veto}}_{n-\mu}=6\tau(^{10}{\rm C})$.} \footnotesize
\begin{tabular*}{86mm}{c@{\extracolsep{\fill}}ccccc}
\toprule & \multicolumn{5}{c} {background index$^a$} \\
\cline{2-6}
 & no & $\mu_{{\rm norm}}$ & \multicolumn{3}{c}{n-associated muon veto} \\
\cline{4-6}
 & veto & veto & 2 $\tau_{^{10}{\rm C}}$ & 4 $\tau_{^{10}{\rm C}}$ & 6 $\tau_{^{10}{\rm C}}$ \\
\hline
efficiency $\varepsilon_\mu$ & 1 & 0.902 & 0.879 & 0.858 & 0.837 \\
$^{10}$C & 16.4 & 14.3 & 1.98 & 0.27 & 0.053 \\[0.4mm]
$^{6}$He & 4.9 & 1.69 & 0.065 & 0.065 & 0.063 \\[0.4mm]
$^{8}$Li & 1.5 & 0.54 & 0.017 & 0.017 & 0.016 \\[0.4mm]
$^{12}$B & 1.9 & 0.05 & 3.8e-4 & 3.8e-4 & 3.8e-4 \\[0.4mm]
others & 0.51 & 0.17 & 0.01 & 0.01 & 0.01 \\
total bkg & 25.2 & 16.8 & 2.1 & 0.36 & 0.14 \\ \bottomrule
\end{tabular*}
\vspace{0mm}
\end{center}
\vspace{0.5mm} \noindent {\footnotesize

$^a$in /ROI/(ton $^{136}$Xe)/yr unit}\vspace{4mm}

\subsubsection{$^{137}$Xe background}

The neutrons that are produced by the cosmic muons can thermalize
via collision with the nuclei in the LS, then finally get captured
on a nuclide. $^{136}$Xe can capture the thermal neutrons and
produce radionuclide $^{137}$Xe via the $^{136}$Xe$({\rm n},
\upgamma)$ process, although the probability is small. $^{137}$Xe
atoms are produced in a capture state with the excited state
energy of 4025.46$\pm$0.27 keV~\cite{Mughabghab}, then de-excite
into the ground state promptly, primarily through $\upgamma$
emission. The ground state of $^{137}$Xe then purely $\upbeta^-$
decays ($\tau=5.51$ min, $Q=4173\pm7$ keV), resulting in
contamination of the ROI.

Similar to $^{10}$C and $^{6}$He, the $^{136}$Xe$({\rm n},
\upgamma)^{137}$Xe cascade also provides a nice triple-coincidence
signature of the muon, the neutron capture on $^{136}$Xe and the
subsequent $^{137}$Xe decay, to identify and reject such
muon-induced $^{137}$Xe background. The neutron capture on
$^{137}$Xe is easy to identify due to a much higher energy than
the natural radioactivity.

The $^{137}$Xe production was estimated from the neutron capture
process in the Xe-LS, as shown in Table~4. The expected neutron
capture fractions on protons, $^{10}$C, $^{136}$Xe and $^{134}$Xe
in the KamLAND-Zen Xe-LS were reported as 0.994, 0.006,
9.5$\times10^{-4}$ and 9.4$\times10^{-5}$,
respectively~\cite{::2015uaa}. In Section~2, we considered doping
5\% by weight of $^{{\rm enr}}$Xe with 80\% $^{136}$Xe into the
JUNO LS, thus the neutron capture fraction on $^{136}$Xe is
expected to be $\sim1.7\times10^{-3}$. Since KamLAND-Zen observed
a $\sim$13\% increase in the spallation neutron flux in the Xe-LS
relative to the normal LS~\cite{KamLANDZen:2012aa}, thus a factor
of 1.13 was taken into account when estimating the neutron rate in
JUNO Xe-LS. In addition, our ROI region is a factor of 4 narrower
than KamLAND-Zen due to better energy resolution, as shown in
Table~1. Finally the background index from $^{137}$Xe was
calculated to be 2.3/ROI/(ton $^{136}$Xe)/yr.

Similar to the $\mu_{{\rm n-assoc}}$ veto, we can develop
$^{137}{\rm Xe}$-associated muon ($\mu_{{\rm Xe}-{\rm assoc}}$)
veto criteria:
\begin{itemize}
  \item
  {\it $\mu_{{\rm Xe}-{\rm assoc}}$} identification: among the $\mu_{{\rm norm}}$ samples, if a n-$^{136}$Xe capture candidate occurs within 1
  ms after a muon and within $(R_{{\rm Xe}}+1)$ meters from the detector center.\\
  {\it $\mu_{{\rm Xe}-{\rm assoc}}$} veto: any signal within 1 meter of each associated n-$^{136}$Xe signature and within a veto time window of $5\tau(\text{$^{137}$Xe})$ was rejected.
\end{itemize}

A FLUKA simulation with the EXO-200 detector showed that thermal
neutron capture was the absolute dominant production process for
$^{137}$Xe~\cite{EXO200::2015wtc}. Our GEANT4 simulation with
Xe-LS gave consistent results. After applying the above veto
scheme to the GEANT4 MC data set, the residual $^{137}$Xe
$\upbeta$-decay was 0.07/ROI/(ton $^{136}$Xe)/yr.

\begin{center}
\tabcaption{The estimated $^{137}$Xe production rate via
$^{136}$Xe$({\rm n}, \upgamma)$ process in the assumed JUNO Xe-LS
detector, which was scaled from the KamLAND-Zen detector.}
\footnotesize
\begin{tabular*}{86mm}{c@{\extracolsep{\fill}}ccc}
\toprule & KamLAND-Zen & JUNO Xe-LS \\
\hline
 $R_n$ in Xe-LS$^a$ & 0.045~\cite{Abe:2009aa} & 0.073~\cite{An:2015jdp}\\
n-$^{136}$Xe fraction & 9.5$\times10^{-4}$~\cite{KamLANDZen:2012aa,::2015uaa} & 1.7$\times10^{-3}$ \\
$^{136}$Xe$({\rm n}, \upgamma)^{137}$Xe yield$^b$ & 61 & 98 \\
background index$^c$& 8.2 & 2.3 \\ \bottomrule
\end{tabular*}
\vspace{0mm}
\end{center}
\vspace{-0.6mm} \noindent {\footnotesize

$^a$in Hz/(kton Xe-LS) unit

$^b$in (ton $^{136}$Xe)$^{-1}\cdot$yr$^{-1}$ unit

$^c$in ROI$^{-1}\cdot$(ton $^{136}$Xe)$^{-1}\cdot$yr$^{-1}$ unit}

\subsection{Background summary}

The ROI backgrounds are summarized in Table~5. We evaluated the
total background index for various Xe-LS target sizes, and with
little difference. Other backgrounds, such as $(\upalpha, {\rm
n})$ reactions, were also evaluated and found to be much less than
the components in Table~5. The reduction of the cosmogenic
backgrounds in each muon veto step is shown in Fig.~1.

\begin{center}
\tabcaption{Summary of the projected backgrounds in the
0$\upnu\upbeta\upbeta$ ROI. For light cosmogenic isotopes, the
values are from GEANT4 MC, while for FLUKA MC the total residual
background would increase 0.07/ROI/(ton $^{136}$Xe)/yr. }
\footnotesize
\begin{tabular*}{73mm}{c@{\extracolsep{\fill}}ccc}
\toprule \multicolumn{2}{c}{summary of backgrounds in 0$\upnu\upbeta\upbeta$ ROI} \\
\multicolumn{2}{c}{[ROI$\cdot$(ton $^{136}$Xe)$\cdot$yr]$^{-1}$} \\
\hline
2$\upnu\upbeta\upbeta$ & 0.2 \\
$^{8}$B solar $\upnu$ & 0.7 \\
\\
\multicolumn{2}{c}{cosmogenic background} \\
\hline
$^{10}$C &  0.053\\
$^{6}$He &  0.063  \\
$^{8}$Li &  0.016  \\
$^{12}$B &  3.8$\times10^{-4}$  \\
others ($Z\leqslant$6) &  0.01  \\
$^{137}$Xe &  0.07  \\
\\
\multicolumn{2}{c}{internal LS radio-purity (10$^{-17}$ g/g)} \\
\hline
$^{214}$Bi ($^{238}$U chain) &  0.003 \\
$^{208}$Tl ($^{232}$Th chain) &  --- \\
$^{212}$Bi ($^{232}$Th chain) &  0.03 \\
\\
\multicolumn{2}{c}{external contamination} \\
\hline
$^{214}$Bi (Rn daughter) & 0.2 \\
\\
total &  1.35 \\ \bottomrule
\end{tabular*}
\vspace{0mm}
\end{center}
\vspace{0mm}

\begin{center}
\includegraphics{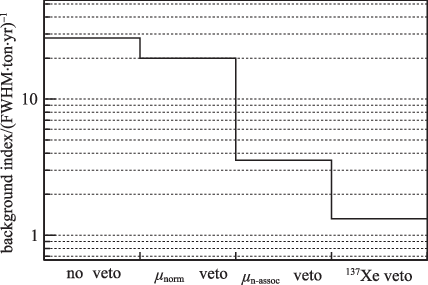}\vspace{1.5mm}
\figcaption{The reduction of the total background index for
different muon veto schemes. The $\mu_{{\rm norm}}$ and $\mu_{{\rm
n-assoc}}$ refer to the normal muon veto and the
neutron-associated muon veto methods, respectively, as described
in Section~3.4.}
\end{center}

\section{Sensitivity}

In an experiment that searches for rare decays, with certain
projected backgrounds and no true signal, the sensitivity $S(b)$
can be given by $S(b) = \sum_{n=0}^{\infty} P(n|b)
U(n|b)$~\cite{GomezCadenas:2010gs}, where $P(n|b)$ is a Poisson
$p.d.f$ for the background fluctuation, and $U(n|b)$ is a function
yielding the upper limit at the desired C.~L. for a given
observation $n$ and a mean projected background level $b$. In a
real $0\upnu\upbeta\upbeta$ experiment with non-negligible
background, the sensitivity of the 0$\upnu\upbeta\upbeta$
half-life can be calculated as
\begin{equation}
  T^{0\upnu\upbeta\upbeta}_{1/2} = \ln{2} \cdot \dfrac{N_A}{M_{{\rm isotope}}}\cdot\dfrac{M\cdot\epsilon\eta\cdot t}{\alpha\cdot\sqrt{b}}
  \label{eqn:sens}
\end{equation}
where $N_A=6.022\times10^{23}$ is Avogadro's constant, $M_{{\rm
isotope}}$ is the molar mass of the $0\upnu\upbeta\upbeta$ decay
isotope, $M$ is the fiducial target mass, $t$ is the live time
(the product $M\cdot t$ is usually referred to as the total
exposure), $\epsilon$ is the detection efficiency, and $\eta$ is
the abundance of the 0$\upnu\upbeta\upbeta$ isotope. In the
calculation, the efficiency $\epsilon$ included the energy cut
efficiency of the 0$\upnu\upbeta\upbeta$ ROI in Table~1 and the
efficiency of the muon veto in Table~3, and the combined
efficiency is listed in Table~6. Depending on whether 90\% or 95\%
C.~L. is quoted, $\alpha$ is 1.64 or 1.96, respectively.

\begin{center}
\includegraphics{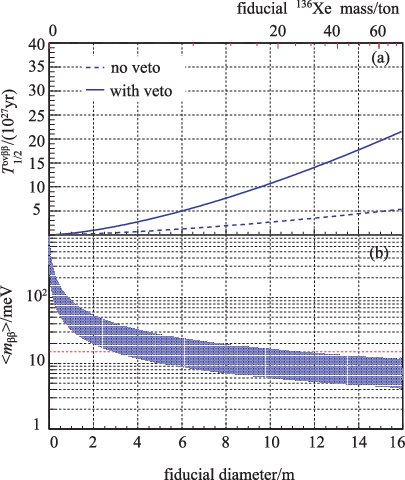}
\figcaption{(color online) (a) The sensitivity of
$T^{0\upnu\upbeta\upbeta}_{1/2}$ versus the JUNO Xe-LS volume size
and fiducial $^{136}$Xe mass assuming 5 years livetime. The two
curves represent the case  w/o and w/ muon veto, respectively, and
the latter is used to calculate the $m_{\upbeta\upbeta}$
sensitivity. (b) The sensitivity of effective neutrino mass
$m_{\upbeta\upbeta}$. The uncertainty band is due to different NME
models (EDF~\cite{Rodriguez:2010mn}, ISM~\cite{Menendez:2008jp},
IBM-2~\cite{Barea:2015kwa}, Skyrme QRPA~\cite{Mustonen:2013zu},
QRPA~\cite{Engel:2014pha}). The red dashed line corresponds to 15
meV.}
\end{center}

Given that the JUNO detector has 20 kton LS, it has the capability
for a large Xe-LS volume. The sensitivities of
$T^{0\upnu\upbeta\upbeta}_{1/2}$ and corresponding effective
neutrino mass $m_{\upbeta\upbeta}$ versus different Xe-LS volume
size and fiducial $^{136}$Xe mass are shown in Fig.~2. The
uncertainty band of $m_{\upbeta\upbeta}$ accounts for different
NME models~\cite{Rodriguez:2010mn,Menendez:2008jp,Barea:2015kwa,
Mustonen:2013zu,Engel:2014pha}. Assuming $\sim$50 tons of
$^\text{136}$Xe and 5 years live time, the projected
$T^{0\upnu\upbeta\upbeta}_{1/2}$ sensitivity (90\% C.L) could
reach $\sim1.8\times10^{28}$ yr with a sophisticated muon veto
scheme, which is a factor of 4 better than the no veto case. This
allows probing of $m_{\upbeta\upbeta}$ down to (5--12) meV,
completely below the region allowed  by the inverted neutrino mass
ordering scenario. We understand that the cost of producing 50
tons enriched xenon is currently practically unacceptable,
however, to demonstrate that the sensitivity could really scale
with target mass, we quote a large target mass.

The sensitivity for a more realistic scenario, with 5 tons of
$^\text{136}$Xe and 5 years live time, is superimposed in Fig.~3.
The projected $T^{0\upnu\upbeta\upbeta}_{1/2}$ sensitivity (90\%
C.L) would be $\sim5.6\times10^{27}$ yr. We also would like to
point out that, since $M$ in Eq.~(2) refers to the fiducial target
mass, the fiducial cut efficiency was automatically included. As
discussed in Section~3.3.2, a 1-m cut from the Xe-LS target edge
was considered, thus the fiducial efficiency was 45\% and 67\% for
5 tons and 50 tons of $^{136}$Xe, respectively. In future work, we
expect to perform a 2D fit simultaneously to the energy spectra
and stand-off distance, which is defined as the distance from the
event position to the Xe-LS edge, in order to enlarge the fiducial
volume and improve the sensitivity.

\end{multicols}
\ruleup
\begin{center}
\tabcaption{A comparison of current and future
0$\upnu\upbeta\upbeta$ experiments, including: the target
0$\upnu\upbeta\upbeta$ isotope and its abundance in the natural
isotopes; the exposure of the 0$\upnu\upbeta\upbeta$ isotope; the
detection efficiency for 0$\upnu\upbeta\upbeta$; the background
index (B.I.); the 90\% C.L. limit or sensitivity of
0$\upnu\upbeta\upbeta$ decay half-life $T^{0\upnu}_{1/2}$; and the
90\% C.L. limit or sensitivity of the effective neutrino mass
$m_{\upbeta\upbeta}$. Unless specially noted, the background
index, in events/(keV ton yr) unit, is defined as the background
counts normalized by the ROI width and the 0$\upnu\upbeta\upbeta$
isotope exposure.} \footnotesize
\begin{tabular*}{175mm}{@{\extracolsep{\fill}}cccccccc}
\toprule experiment & isotope & exposure & $\varepsilon_{0\upnu\upbeta\upbeta}$ & B.I. & ROI & \multicolumn{2}{c}{90\% C.L. limit (L) or sensitivity (S)}\\
\cline{7-8}
 &  & /(ton$\cdot$yr) &  &  & /keV & $T^{0\upnu}_{1/2}$, $\times10^{27}$yr & $m_{\upbeta\upbeta}$/meV  \\
\hline
\multicolumn{8}{c}{current results}\\
\hline
CUORE-0~\cite{Alfonso:2015wka} & $^{130}$Te (34.17\%) & 9.8e-3 & 0.813 &  58$^a$ & 5.1 $^\text{FWHM}$ & 0.004$^b$ (L) & $270-760$ (L) \\
EXO-200~\cite{Albert:2014awa} & $^{136}$Xe (80.6\%) & 0.1 & 0.846 & 1.7$^c$ & 150 (2$\sigma$) & 0.019 (S) & $190-450$ (L) \\
\cline{3-6} GERDA~\cite{Agostini:2016} & $^{76}$Ge (87\%) &
5e-3$^d$ & 0.51 & 3.5 & 10.2 (3$\sigma$) & 0.04$^e$ (S) & $160-260$ (L)  \\
(phase-II)& & 5.8e-3 & 0.60 & 0.7 & 7.7 (3$\sigma$) &  & \\
\cline{3-6} KamLAND-Zen~\cite{KamLAND-Zen:2016pfg} & $^{136}$Xe
(90.77\%) & $\sim$0.255  & --- & 28.1/yr & 400 & 0.056 (S), 0.092
(L) & $61-165$ (L)$^f$  \\
(phase-II) &  & $\sim$0.249  & --- & 15.5/yr & 400 & &   \\
\cline{3-6}
\\
\multicolumn{8}{c}{prospective sensitivities}\\
\hline
EXO-200 phase-II~\cite{Yang2016} & $^{136}$Xe & $\sim$0.16 $\cdot$ 3 & --- & --- & ---& 0.057 (S) & $110 -260$ (S)\\
KamLAND-Zen 800~\cite{KamLAND-Zen:2016pfg} & $^{136}$Xe & $\sim$0.8 $\cdot$ ?  & --- & --- & --- & --- &  $\sim$50 (S)\\
SNO+ phase I~\cite{Andringa:2015tza} & $^{130}$Te & $\sim$0.8 $\cdot$ 5  & --- & 13.4/yr & --- & 0.09 (S) & $55-133$ (S)\\
CUORE~\cite{Canonica2016} & $^{130}$Te & 0.206 $\cdot$ ? & --- & 10 & --- & 0.095 (S) & $50 -130$ (S) \\
GERDA Phase-II~\cite{Agostini:2016} & $^{76}$Ge & $>$0.1  & --- & $\sim$1 & --- & $>$0.1 (S) & --- \\
\\
SNO+ Phase II~\cite{Andringa:2015tza} & $^{130}$Te & $\sim$8.0 $\cdot$ ?  & --- & --- & --- &  0.7 (S) & $19-46$ (S) \\
KamLAND2-Zen~\cite{Shirai2016} & $^{136}$Xe & $\sim$1 $\cdot$ ?  & --- & --- & --- & --- & $\sim$20 (S)  \\
nEXO~\cite{Lin2015} & $^{136}$Xe (90\%) & $\sim$5 $\cdot$ 5 & --- & 0.02$^g$ & 58$^{\text{FWHM}}$ & 6.6 (S) & $7-22$ (S) \\
JUNO Xe-LS & $^{136}$Xe (80\%) &  50 $\cdot$ 5 & 0.63 & 0.012 & 110$^{\text{FWHM}}$ & 18 (S) & $5-12$ (S) \\
\bottomrule
\end{tabular*}%
\end{center}
\vspace{1mm} \noindent {\footnotesize

$^a$The quoted B.I. is normalized to the total TeO$_{2}$ exposure
(35.2 kg yr). The same for CUORE.

$^b$This limit is from the combination with the 19.75 kg$\cdot$yr
exposure of $^{130}$Te from Cuoricino, while it is
$2.7\times10^{24}$ yr for CUORE-0 only.

$^c$This quoted B.I. is normalized to the total Xe exposure (123.7
kg yr).

$^d$The quoted 5 (5.8) kg$\cdot$yr exposure is for the total
coaxial (BEGe) detectors in GERDA Phase-II.

$^e$The limits of $T^{0\upnu}_{1/2}$ and $m_{\upbeta\upbeta}$ are
from the combination of Phase-I and Phase-II. For Phase-I only, it
was $2.1\times10^{25}$ yr (90\% C.~L.).

$^f$The quoted limit is from the combination of KamLAND-Zen
Phase-I and Phase-II.

$^g$The quoted B.I. is for the inner 3-ton xenon mass.}\\

\ruledown
\begin{multicols}{2}

\section{Summary and discussion}

In this work, we explored the physics potential of a
0$\upnu\upbeta\upbeta$ search with the JUNO detector via
dissolving $^{136}$Xe-enriched xenon gas into LS. JUNO is designed
to achieve 3\%/$\sqrt{E \text{(MeV)}}$ energy resolution for
determining neutrino mass ordering, thus the energy resolution at
$^{136}$Xe $Q_{\upbeta\upbeta}$ is expected to be 1.9\%, resulting
in relatively small intrinsic 2$\upnu\upbeta\upbeta$ background in
the ROI. We performed detailed analyses of other ROI backgrounds
from $^8$B solar $\upnu$-e scattering events, LS natural
radioactivity, muon induced radionuclides, and so on. An optimal
purity (10$^{-17}$ g/g for $^{238}$U and $^{232}$Th) is assumed
with proper LS on-line purification. A sophisticated muon veto
scheme using the correlation between the spallation neutrons and
the isotopes was developed to reject the long-lived cosmogenic
backgrounds. Eventually a low background rate of
$\sim$1.35/ROI/(ton $^{136}$Xe)/yr was expected to be achievable.
Assuming 5 tons of fiducial $^{136}$Xe target mass and 5 years
live time, we projected the 90\% C.L sensitivity of
$T^{0\upnu\upbeta\upbeta}_{1/2}$ (or $m_{\upbeta\upbeta}$) to be
$\sim5.6\times10^{27}$ yr (or 8--22 meV). In the case of 50 tons
of fiducial $^{136}$Xe, the 90\% C.L sensitivity of
$m_{\upbeta\upbeta}$ can scale up to (5--12 meV), which is well
below the region allowed by the scenario of inverted neutrino mass
ordering.

Ultra-low background and excellent energy resolution are the two
critical factors for the next generation 0$\upnu\upbeta\upbeta$
experiments. Table~6 summarizes the current experimental results
or the projected sensitivities of
CUORE~\cite{Alfonso:2015wka,Canonica2016},
EXO-200~\cite{Albert:2014awa}, GERDA~\cite{Agostini:2016},
KamLAND-Zen~\cite{KamLAND-Zen:2016pfg},
SNO+~\cite{Andringa:2015tza}, nEXO~\cite{Lin2015,Yang2016}, as
well as the potential Xe-LS detector at JUNO. Different
experiments use different definitions when reporting the
background rate, as well as choosing different
0$\upnu\upbeta\upbeta$ windows. In order to compare different
experiments, we rewrite the sensitivity formula Eq.~2
as\vspace{1.5mm}
\begin{equation}
  \left(\dfrac{T^{0\upnu\upbeta\upbeta}_{1/2}\cdot\alpha}{\ln{2}\cdot N_A}\right)^2 = \dfrac{M_{{\rm norm}}}{B_{\rm I}}\vspace{2mm}
  \label{eqn:sensNew}
\end{equation}
where $B_{\rm I}=\dfrac{b}{(M\epsilon\eta\cdot t/M_{{\rm
isotope}})\cdot \text{ROI}}$ is the \vspace{1.5mm}redefined
background index, and $M_{{\rm norm}}=\dfrac{M\epsilon\eta\cdot
t}{\text{ROI}\cdot M_{{\rm isotope}}}$ is the normalized detector
exposure.

With the new definition, Fig.~3 shows a comparison of the
experiments listed in Table~6. The dashed lines represent the
contours of different sensitivities of
$T^{0\upnu\upbeta\upbeta}_{1/2}$ using Eq.~(3). The data points
roughly agree but do not exactly align with the calculated
contours, because different experiments have different systematics
and use different fitting or statistical analysis methods. Fig.~3
also indicates that the next generation 0$\upnu\upbeta\upbeta$
experiments should pursue both ultra-low background and very large
detector exposure.

\vspace{1.5mm}
\begin{center}
\includegraphics{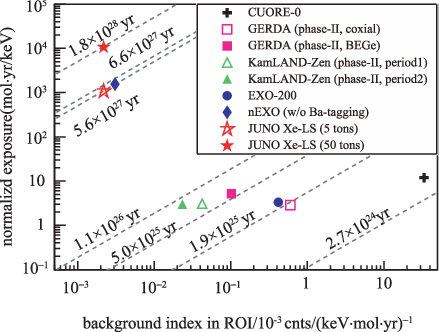}\vspace{1.5mm}
\figcaption{(color online) The re-defined background indices and
detector exposures in the ROIs of CUORE-0~\cite{Alfonso:2015wka},
GERDA Phase-II~\cite{Agostini:2016},
KamLAND-Zen~\cite{KamLAND-Zen:2016pfg},
EXO-200~\cite{Albert:2014awa}, nEXO~\cite{Yang2016} and the
potential Xe-LS detector at JUNO. The dashed lines are the
contours of different sensitivities.}
\end{center}

\end{multicols}

\vspace{-2.5mm} \centerline{\rule{80mm}{0.1pt}} \vspace{1mm}

\begin{multicols}{2}

\end{multicols}

\clearpage


\end{document}